\documentstyle[12pt]{article}

\newcommand{\tr}{\mbox{tr}}
\newcommand{\Od}{{\cal O}}
\newcommand{\Ima}{\hbox{Im}}

\newcommand{\NP}[1]{{\em Nucl.\ Phys.\ }{\bf #1}}
\newcommand{\ZP}[1]{{\em Z.\ Phys.\ }{\bf #1}}

\newcommand{\PL}[1]{{\em Phys.\ Lett.\ }{\bf #1}}

\newcommand{\PR}[1]{{\em Phys.\ Rev.\ }{\bf #1}}

\def\gappeq{\mathrel{ \rlap{\raise.5ex\hbox{$>$}}
                      {\lower.5ex\hbox{$\sim$}}  } }
\def\lappeq{\mathrel{ \rlap{\raise.5ex\hbox{$<$}}
                      {\lower.5ex\hbox{$\sim$}}  } }
\begin{document}
\input epsf
\thispagestyle{empty}
\hfill    LBNL-39378

\hfill    September 1996

\begin{center}
{\LARGE \bf  Resonance Spectrum\\
of the \\
strongly interacting \\
Symmetry Breaking Sector
}
\footnote{This work was supported by the Director, Office of Energy
Research, Office of High Energy and Nuclear Physics, Division of High
Energy Physics of the U.S. Department of Energy under Contract
DE-AC03-76SF00098.}

\vskip 1.cm

\Large J.R. Pel\'aez\normalsize$^{2}$ \\
Theoretical Physics Group. Lawrence Berkeley Laboratory\\
University of California.
Berkeley, California 94720. USA.

\end{center}

\vskip .5cm

\begin{abstract}
Within the chiral lagrangian formalism it is
possible to describe the general
strongly coupled Symmetry Breaking Sector 
in terms of a few parameters.
Based on a dispersive approach we have
studied the resonance spectrum up to 3 TeV
in the chiral parameter space.
This procedure could also be useful
to extract the higher energy resonant behavior
from low-energy collider data.
It is also shown how the method reproduces the 
correct pole structure
of resonances as well as other analytic features.
The results also hint at a possible excluded
region of parameter space.
\end{abstract}

\footnotetext[2]{
Complutense del Amo fellow. On leave of absence from:
Departamento de F\'{\i}sica Te\'orica. Universidad Complutense.
28040 Madrid, Spain. E-mail:pelaez@theorm.lbl.gov, pelaez@vxcern.cern.ch}
\newpage

\section{Introduction}

The main purpose of the next generation of colliders
is to unveil the nature of the electroweak Symmetry Breaking Sector
(EWSBS). Despite the remarkable success of the 
Standard Model (SM) with the present
precision electroweak data, the mechanism
responsible of this breaking remains unknown. There are, however, many 
theoretical models which, very roughly, 
can be divided in two
categories: weakly or strongly coupled.

In the weak case light particles are expected
below the TeV scale. Typical examples are the Minimal SM
(MSM) with a light Higgs or most supersymmetric models.
These models have become very popular and have been 
studied in great detail. That is not possible
in the strong case, where 
the strength of the interactions makes
the usual perturbative approach unreliable.
In particular, there are no light 
particles to control the generic
enhancement of gauge boson production. As a consequence, 
the perturbative calculations suffer from
severe unitarity violations.
Nevertheless, such an enhancement would be the experimental
signature of an strong EWSBS. The most promising process 
is longitudinal gauge boson scattering, where
the most striking feature
would be the appearance of heavy resonances.

There are also several models of strongly coupled
EWSBS, like the MSM with a heavy Higgs boson, Technicolor, 
composite models, etc... From very general symmetry considerations
all them share the same dynamics at low energies \cite{LET}. 
However, the predictions of these models can vary 
greatly from one another.
 
Several years ago it was introduced 
a theoretical framework that is able to describe
generically the strong interactions of electroweak gauge 
bosons \cite{DoHe}.
It is based on Chiral Perturbation Theory (ChPT)
\cite{GaLe}, which works 
remarkably well for pion physics.
The idea is to write an effective chiral lagrangian 
including operators up to dimension four \cite{Appel}.
The form of the terms is only constrained 
by symmetry considerations which are common to
any strong EWSBS. Thus, using this lagrangian it is possible
to mimic the {\em low energy behavior} of any strong EWSBS.
The difference between underlying
theories appears through the values of the parameters in the 
chiral lagrangian. There are already published
chiral parameter estimates for several models like 
the MSM with a heavy Higgs \cite{HeRu,EsMa} or Technicolor
\cite{TC}. There are also studies 
which indicate that at least part of the interesting 
parameter space will be accessible at LHC \cite{CMS}.

However, the usual chiral lagrangian approach does
not respect unitarity. At low energies the violations are
very small, but they increase with the energy. 
As a consequence it is not able to reproduce resonances
unless it is modified. There are several ways
to unitarize chiral amplitudes. Many of them are simple 
mathematical tricks whose results 
very frequently differ, which is an obvious criticism to such 
procedures.  Nevertheless, over the last  
few years, it has been developed a technique, known as the Inverse
Amplitude Method (IAM) \cite{Truong,Pade1,Pade2,IAM},
whose results have been successfully tested in ChPT.
It is based on dispersion theory and it can accommodate
all the analytic structure required to reproduce resonances.
Namely, the elastic cut and poles in the second Riemann sheet.
When applied to low energy hadron physics, it is indeed able to
reproduce the lightest resonances. The IAM
seems very reliable at least at the qualitative level.

Concerning the EWSBS, the method was first applied
to mimic a heavy Higgs and a QCD-like scenario at 
supercolliders \cite{DoHeTe}. 
The results of the IAM are once again consistent
with the expected resonances.

The aim of this work is to explore
the interesting part of the chiral parameter space 
using the IAM method. In so doing, we expect to
obtain a description of the low resonance spectrum
of the general strong EWSBS.

The paper is organized as follows: In section 2 we 
discuss the theoretical framework used in this work. 
First we 
introduce to the chiral formalism; next,
we address some technical issues related to the Equivalence Theorem.
We then define partial waves and state the unitarity problem.
As a solution, it is briefly reviewed the IAM, whose derivation
is given in the Appendix.
Section 3 is devoted to the IAM results.
First for reference models, that we use to illustrate
different analytical and physical features, like saturation.
We then show where these phenomena appear in parameter space.
The problem of unitarity in the
$I=2$ channel and whether it can be used to exclude part of the
parameter space is also addressed in Section 3.
In section 4 we discuss these results and 
we gather them in the conclusion.

\section{Resonances in the chiral formalism}

\subsection{The chiral lagrangian}

Let us remember that we have to break the $SU(2)_L\times U(1)_Y$
{\em gauge} symmetry down to $U(1)_{EM}$. 
Therefore we need a {\em global} breaking from a group $G$ down 
to another $H$. It should provide three Goldstone Bosons
(GB) that will become the logitudinal components of the gauge bosons
through the Higgs mechanism. We also want to include the
custodial $SU(2)_{L+R}$, which naturally yields a $\rho\simeq1$
parameter \cite{Sikivie}. It can be shown 
that these constraints lead to 
$G=SU(2)_L\times SU(2)_R$ and $H=SU(2)_{L+R}$ \cite{LET,NPB}.
Thus, the GB fields $\pi^i$ can be seen as coordinates in the
$G/H\sim SU(2)_{L-R}$ coset. Hence, we will parametrize them in an
$SU(2)$ matrix as $U=\exp(i\pi^i \sigma^i/v)$. The parameter 
$v\simeq256\mbox{GeV}$  plays here the
same role as $f_\pi$ in ChPT and sets the scale of the
EWSBS.

Within the chiral approach we build the low-energy lagrangian
as an expansion in derivatives (momenta) of the GB fields.
Since we will work up to $\Od(p^4)$, we 
should look for a complete set of $SU(2)_L\times U(1)_Y$,
Lorentz, C and P invariant operators containing up to four derivatives. 
These have been obtained
in \cite{Appel}, but they are too general for our purposes.
Indeed, we want an exact 
$SU(2)_{L+R}$ symmetry on the hidden sector
once the gauge couplings, $g$ and $g'$,
are set to zero.
In addition, we are only interested in gauge boson 
elastic scattering and we can neglect $CP$ violating effects.

With those assumptions, the only operator that 
we can build with two derivatives is

\begin{equation}
{\cal L}^{(2)}=\frac{v^2}{4}\tr D_\mu UD^\mu U^\dagger
\label{NLSM}
\end{equation}
where
$D_\mu U= \partial_\mu U-W_\mu U+UB_\mu$ is a covariant derivative
with $W_\mu=-ig\sigma^aW^a_\mu/2$ and  
$B_\mu=-ig\sigma^3B_\mu/2$.
It is important to observe that this lagrangian only depends 
on the symmetry breaking
pattern and the scale.
In this sense, the amplitudes obtained from 
${\cal L}^{(2)}$ are universal. That is why they are called 
Low Energy Theorems (LET) \cite{LET}. 

Notice also that the lagrangian in Eq.\ref{NLSM}
is that of the non-linear
$\sigma$ model and thus it is not renormalizable.
In fact it is not possible to absorb 
the loop divergencies by introducing a finite 
set of new counterterms and constants.
Nevertheless, we are only interested in
the low-energy behavior and therefore it is enough to
work up to a given order in the external momenta.
For instance, if we want to obtain gauge boson scattering
amplitudes at $\Od(p^2)$, the only contributions come from 
${\cal L}^{(2)}$ at tree level.
If we calculate at $\Od(p^4)$, we will have to consider
the ${\cal L}^{(4)}$ lagrangian at tree level as well as 
${\cal L}^{(2)}$ to one loop. These last contributions are
divergent, but their divergencies can be absorbed in 
the ${\cal L}^{(4)}$ parameters. In this sense,
the calculations are renormalizable and finite.
This procedure can be generalized to $\Od(p^N)$,
but we will work only up to $\Od(p^4)$.

There are many possible terms in the ${\cal L}^{(4)}$ lagrangian
\cite{Appel}. However, according to the above restrictions,
we are only interested in
\begin{eqnarray}
{\cal L}^{(4)}&=&
L_1 \left( \tr D_\mu UD^\mu U^\dagger \right)^2
+ L_2 \left( \tr D_\mu UD^\nu U^\dagger \right)^2\nonumber \\
&+& \tr \left[
(L_{9L} W^{\mu\nu}+L_{9R} B^{\mu\nu})
D_\mu UD_\nu U^\dagger
\right]
+ L_{10}\tr U^\dagger B^{\mu\nu} U W_{\mu\nu}
\label{L4}
\end{eqnarray}
where $W^{\mu\nu}$ and $B^{\mu\nu}$ are the strength tensors
of the gauge fields.

Finally, let us remark
that using these lagrangians we will obtain the chiral amplitudes as
truncated series in $s$, the usual Mandelstam variable. That is
\begin{equation}
t(s)\simeq t^{(0)}(s) + t^{(1)}(s) + \Od(s^3)
\label{trunc}
\end{equation}
Where $t^{(0)}(s)$ is $\Od(s)$ and reproduces the LET.
It is obtained from ${\cal L}^{(2)}$ at tree level.
The $t^{(1)}(s)$ contribution is $\Od(s^2)$ and
comes from the ${\cal L}^{(4)}$ at tree level and
${\cal L}^{(2)}$ at one loop.
The loops yield logarithmic contributions
which are very relevant at low energies. However, at higher
energies our amplitudes behave essentially as polynomials in $s$.

\subsection{Chiral parameters}

In contrast with the ${\cal L}^{(2)}$ lagrangian, the one
in Eq.\ref{L4} is not completely fixed by symmetry
 and the scale. Indeed each operator has
a parameter which depends on the specific breaking
mechanism. Thus, for every strong EWSBS 
without relevant light modes and our assumed
symmetry breaking pattern,
there should be a different set of chiral parameters.
Notice, however, that nothing ensures the reciprocal.
It is not clear that for every set of chiral parameters there should be 
an underlying consistent and renormalizable Quantum Field Theory (QFT). 

Unfortunately, the very nature of strongly coupled theories
does not allow a calculation of these parameters. There are, however,
estimates for the 
heavy Higgs MSM, which are obtained from a matching
of one loop Green functions \cite{HeRu}. For the QCD-like model, they are
obtained by rescaling the QCD parameters \cite{IAM}. 
We will use these models as a reference and thus we have listed 
their parameters in Table 1.

\begin{center}
\begin{tabular}{|l|cc|}\hline
& $L_1$ & $L_2$\\ \hline
MSM  ($M_H\sim1$ TeV)& 0.007 & -0.002 \\
QCD-like & -0.001 & 0.001 \\ \hline
\end{tabular}

\vskip 3mm

{\footnotesize {\bf Table 1: }
Chiral Parameters for different reference models.}
\end{center}

Very recently several studies have appeared
concerning the LHC capabilities to determine 
these parameters in case there is a strong EWSBS \cite{CMS}.
Notice that their expected values  are in the $10^{-2}$ to $10^{-3}$
range.
Note also that the sign of the parameters may play an essential role.
From these preliminary studies it seems that LHC could
be able to reach the $5\times10^{-3}$, even in the hardest 
non-resonant case. However that will require two detectors
taking data for several years
at full design luminosity and the highest center of
mass energy.

\subsection{The Equivalence Theorem}

As we have already seen, the most relevant modes of the
EWSBS at low energy are the GB. However once we include the electroweak
interactions, the GB disappear from the physical spectrum 
and become the longitudinal components of the
gauge bosons ($V_L$).
 Somehow we can identify the GB and their behavior with that
of the gauge bosons. The precise formulation of the previous statement is known
as the Equivalence Theorem (ET) \cite{ET1,ChaGa}:
\begin{equation}
T(V^{a_1}_L,...V^{a_n}_L)\simeq 
\left( \prod_{j=1}^{l}K^{a_j}_{\alpha_j} \right)
T(\pi_{\alpha_{1}} ...\pi_{\alpha_n})+\Od\left(\frac{m}{E}\right)
\end{equation}
where $m$ is the mass of the gauge boson. 
The $K$ factors, which include renormalization and higher
order $g$ effects, are  basically
$1+\Od(g^2)$ \cite{ET2}. In short, the ET allows us to 
identify, at energies $E\gg m$, the longitudinal gauge boson
amplitudes with those of their associated GB.
It is very useful in two senses: First it allows to link 
the physical measurements with the hidden sector.
Second, it helps in the calculation of the $V_L$ amplitudes,
which are much easier to obtain using scalar particles like GB.

It is important to notice that the ET is a high energy limit.
In contrast, the chiral formalism is a low energy approach.
Nevertheless, it has been recently shown that there
is a window of applicability for the ET together with the
chiral approach \cite{NPB,chinos}. The above equation remains valid, but 
only at {\em lowest order} in g and g'.

In the following sections we will be using thoroughly the ET.
Therefore we will work at
lowest order in the electroweak couplings.
As a consequence, only $L_1$ and $L_2$ 
will be relevant for our
calculations.

\subsection{Partial waves, unitarity and Resonances}

As far as we have an $SU(2)_{L+R}$ symmetry in
the EWSBS, we can also define a weak isospin $I$. In analogy
to $\pi\pi$ scattering, we have three possible weak isospin
channels $I=0,1,2$. It is then usual to project the amplitudes
in partial waves with definite angular momentum $J$ and isospin $I$.
At low energies we are only interested in
the lowest $J$, and thus we will study the
$t_{IJ}=t_{00},t_{11}$ and $t_{20}$ partial waves.
Their expressions for the EWSBS where given in \cite{DoHeTe}. 
Customarily the results of elastic scattering are presented  in
terms of their complex phases, which are known as phase shifts.

As we have already remarked, one of the most striking features
of an strongly interacting EWSBS could be the appearance of resonances.
For instance, for the MSM
with $M_H\simeq1\mbox{TeV}$, we expect a
very broad scalar resonance around 1 TeV. In QCD-like models one
expects a vector resonance (similar to the $\rho$ in pion physics)
around 2 TeV.

However, the chiral formalism by itself is not able to reproduce resonances.
Their very existence is closely related to the
saturation of unitarity. But the chiral amplitudes do not even
satisfy the elastic unitarity condition
\begin{equation}
\mbox{Im} t_{IJ}(s) = \sigma(s) \vert t_{IJ}(s) \vert^2
\label{uni}
\end{equation}
where $\sigma(s)$ is the two body phase-space.
Nevertheless, they satisfy it perturbatively
\begin{equation}
\mbox{Im} t^{(1)}_{IJ}(s) = \sigma(s) \vert t^{(0)}_{IJ}(s) \vert^2
+ \Od(s^3)
\label{pertuni}
\end{equation}
Notice that the violation of unitarity is very small 
only at low energies.

Therefore, in order to accommodate resonances we have to unitarize the 
chiral amplitudes. There are many mathematical tricks to impose
unitarity, which very often lead to different results.
Obviously, that is the main criticism to
unitarization. There is, however, a method that has been tested in ChPT
and is able to reproduce the $\rho$ and
$K^*$ resonances \cite{Pade1,Pade2,IAM}. It is based on dispersion theory
and apart from satisfying Eq.\ref{uni}, it
provides the correct unitarity cut on the complex $s$ plane, as
well as poles in the second Riemann sheet.

\subsection{The Inverse Amplitude Method}

Elastic amplitudes in the complex $s$ plane have
a left and a right (or unitarity) cut. A dispersion relation
is nothing but Cauchy's Theorem applied to these amplitudes.
They are very useful since we can obtain the values
of the amplitude in any point in terms of integrals
of their imaginary parts over the cuts.

We have just seen that chiral amplitudes 
are not a good approximation at high energies
on the elastic cut. Thus, they are not very well suited 
for a dispersive approach. The key point is to notice that
we can calculate the imaginary part of the inverse amplitude
{\em exactly} on the {\em elastic cut}. Indeed, using Eqs.\ref{uni}
and \ref{pertuni}
\begin{equation}
\Ima \frac{1}{t_{IJ}}
=-\frac{\Ima t_{IJ}}{\mid t_{IJ}\mid^2}=
-\sigma 
\label{ImG}
\end{equation}
We can thus write a dispersion relation for $1/t_{IJ}$ whose
integral over the elastic cut is exact. 
Nevertheless, the other analytical features are still approximate.
In the Appendix, we give a detailed derivation and we comment on these
approximations. Finally, it is possible to solve for $t_{IJ}$ and
we get
\begin{equation}
t_{IJ}\simeq
\frac{t_{IJ}^{(0)2}}{
t_{IJ}^{(0)}-t_{IJ}^{(1)} }
\label{IAM}
\end{equation}
That is the IAM. Apart from its simplicity, it has several advantages:
\begin{itemize}
\item At low energies it reduces again to the very same chiral amplitudes
in Eq.\ref{trunc}.
\item It satisfies elastic unitarity, Eq.\ref{uni}, exactly.
\item The right cut is correctly reproduced and we get the appropriate analytic
structure. In particular we get those poles in the $2^{nd}$ Riemann sheet 
which are near the unitarity cut \cite{IAM}.
\item It can be easily extended to higher orders \cite{Pade2,IAM}.
\end{itemize}
Of course it also has limitations. 
We comment them thoroughly in the Appendix. However,
they are mostly related to analytical structures (like some poles
or the left cut), which are far away from the energy range where
we expect the resonances or unitarity effects. In the elastic
region we expect the IAM to be a good approximation.

Indeed, the IAM has been applied both to
pion elastic scattering and $\pi K$ scattering. The first example is
very similar to the EWSBS, although there the GB are massive. 
In both cases it is possible to reproduce the lowest lying resonances:
The $\rho(770)$ and the $K^*(892)$ respectively \cite{Pade1,Pade2,IAM}.
When only low energy data is used, their masses lie about
15\% off from the actual values. It is however possible to
fit the masses and widths using high energy data. Notice that this can be
achieved without introducing any other field or parameter.

It is also important to remark that the IAM also improves considerably
the nonresonant channels \cite{Pade2,IAM}. In fact, the $I=2$ channel
in $\pi\pi$ and $I=3/2$ in $\pi K$ scattering do not present any low
resonance. In spite of that,
the results of the chiral amplitudes only match the data at 
low energies. The IAM results fit the data remarkably well up to much higher
energies.

In addition the appearance
of resonances is completely consistent with the QFT description. They
are associated to poles in the second Riemann sheet, whose position 
is correctly related to the physical mass and width \ref{IAM}.
The analytical structure of the IAM amplitudes is the correct one
in the elastic region. 

We therefore consider that the chiral formalism, together with the IAM,
is a reliable method to obtain, at least, a qualitative description of 
the resonance spectrum in strongly coupled systems.
\nopagebreak
\section{Results}
\subsection{Reference models. Resonances and saturation.}
The IAM in the chiral lagrangian context was first applied to 
a MSM and a QCD-like model in \cite{DoHeTe}. There it was shown that it is
able to reproduce the expected  resonances: 
a broad scalar resonance
in the heavy Higgs MSM, and a Technirho at
about $2\mbox{TeV}$ in the QCD-like model. As an illustration,
we show in Figure 1 the phase shifts 
obtained when the IAM is applied to the chiral amplitudes.

\leavevmode
\begin{center}
\hspace*{-2mm}
\mbox{\epsfysize=6.cm\epsffile{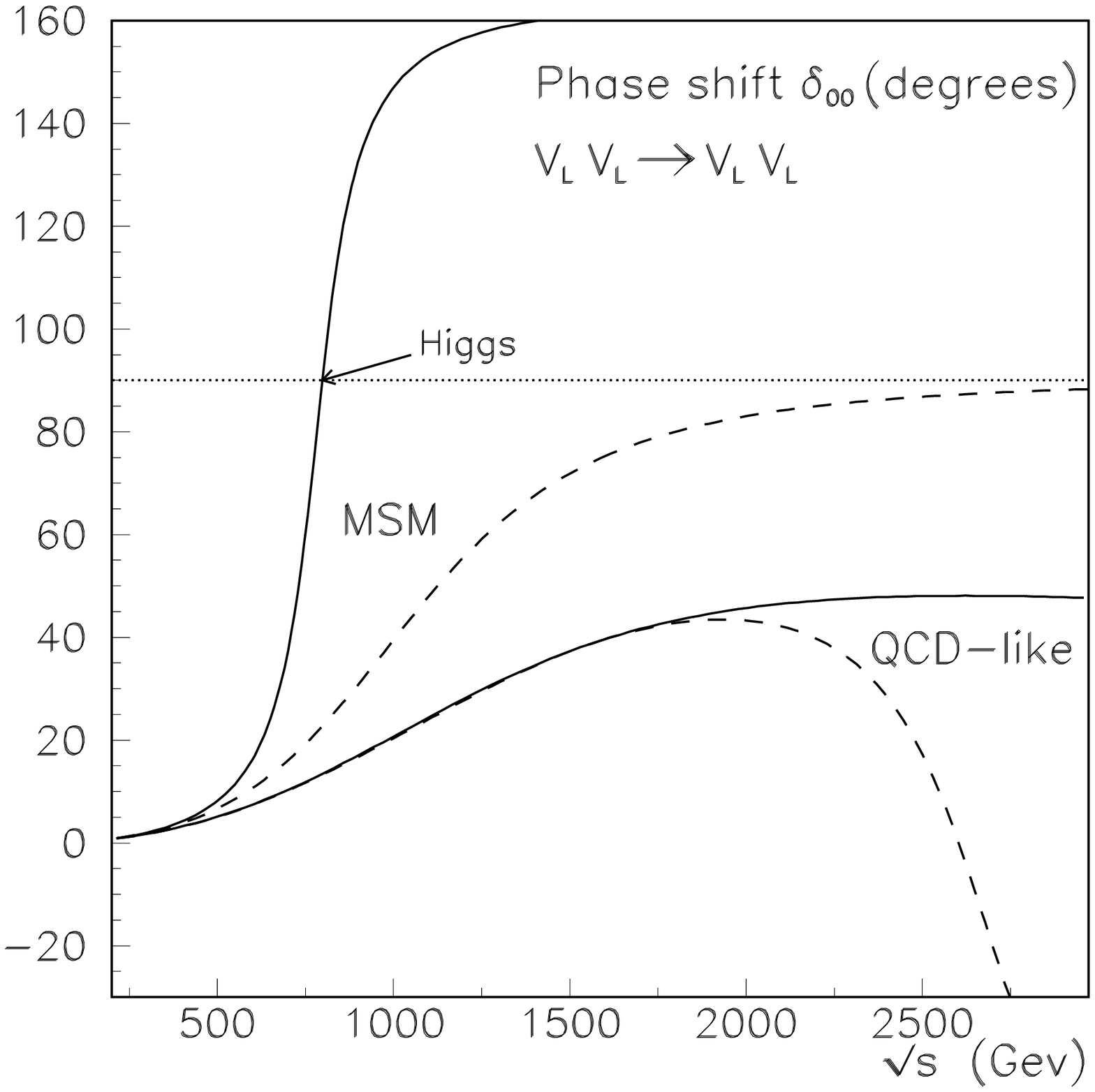}
\hspace{-6mm}
\epsfysize=6.cm\epsffile{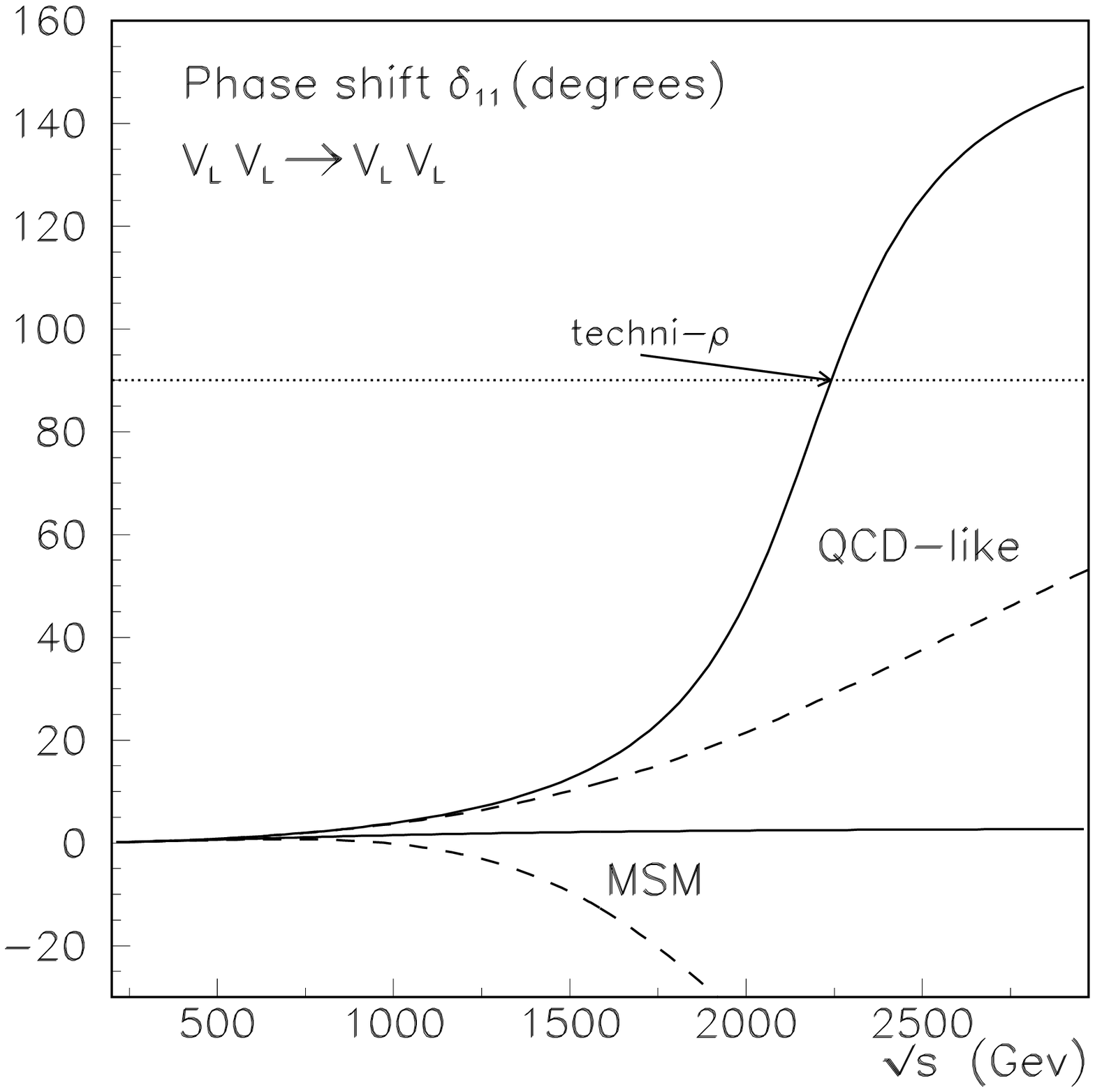}
\hspace{-6mm}
\epsfysize=6.cm\epsffile{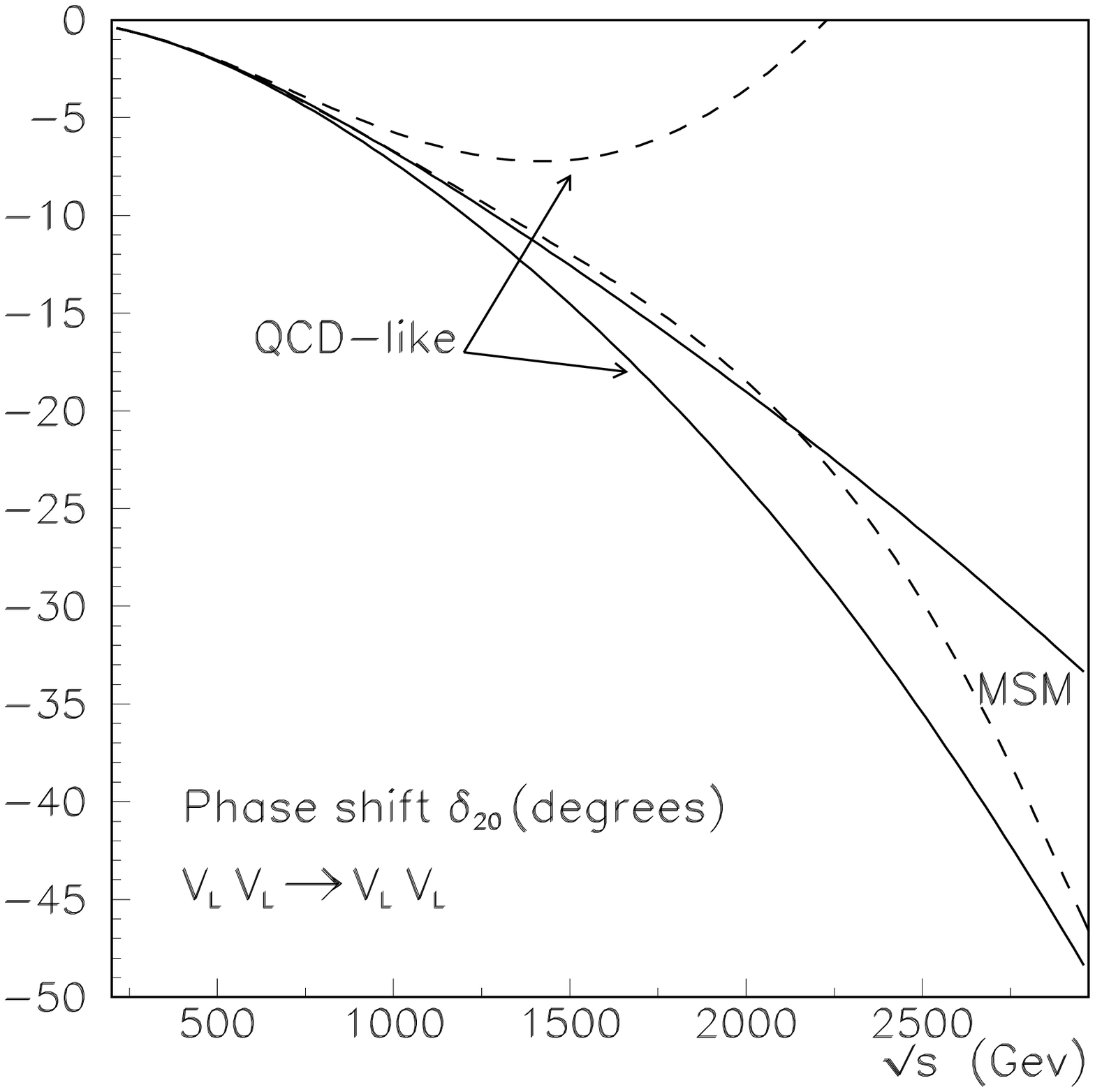}}
\end{center}
\vspace{-.5cm}
{\footnotesize
{\bf Figure 1.-}
Phase shifts in strong $V_LV_L$ scattering.
The dashed lines are the plain chiral amplitudes
and the continuous lines those using the IAM.
They have been obtained both for a heavy-Higgs MSM
and a QCD-like model using the parameters in Table 1.}
\vspace{.5cm}

They have been obtained using the parameters given in 
Table 1, which have been actualized.
Naively, the resonant masses can be obtained from
the point where the corresponding phase shift crosses $90^o$.
The width can be obtained assuming the typical Breit-Wigner
for the Higgs-like resonance and 
$M_\rho\simeq2240\mbox{GeV}$,
$\Gamma_\rho\simeq620\mbox{GeV}$
shape. Their values are: $M_s\simeq800\mbox{GeV}$,
$\Gamma_s\simeq185\mbox{GeV}$ 
for the $\rho$-like resonance.
Notice that in this work we are also giving
the results for the $I=2$ channel. 
It is related to like-sign pair production 
of gauge bosons, where the signal to background ratio seems very
favorable, has it has been pointed out in \cite{ChaGa,likesign}. 
In Figure 1 it can be seen that the results 
using the IAM may vary significantly
from those without unitarization. For instance, in the QCD-like case
even the qualitative behavior is completely different.
Comparing with QCD data, the correct behavior is the one given by the
IAM \cite{Pade2,IAM}.

Finally, in Figure 2, we show the position of the poles in the $2^{\rm nd}$
 Riemann sheet. Figure 2.a is the  pole that appears in the
$(I,J)=(0,0)$ channel when using the MSM parameters of Table 1.
Figure 2.b is the one that appears in the vector channel
when using the QCD-like parameters.
Notice that the positions of the poles satisfy
$\sqrt{s_{pole}}\sim M_{res}+i\Gamma_{res}/2$. 

\subsection{The scalar and vector channels}

\subsubsection{Saturation}

We have been paying an special attention to
resonances, but there are other interesting features. 
In particular, it could happen that the amplitude
saturates unitarity although there is no clear resonant shape.
At this point is important to notice that the 
criterion of $\delta_{IJ}$ crossing $90^o$ is only 
applicable to the cleanest cases. A resonance should be associated
with 
a pole near the real axis
which causes a steep raise in the phase shift.
This pole reflects the existence of an almost bound state.
When there is no other phase background this leads to our naive
$90^o$ criterion. In such cases we can apply the usual Breit-Wigner description
and relate, as above, the resonance physical constants with the pole position.

%-----figure 2
\leavevmode
\begin{center}
\mbox{\epsfysize=6.cm\epsffile{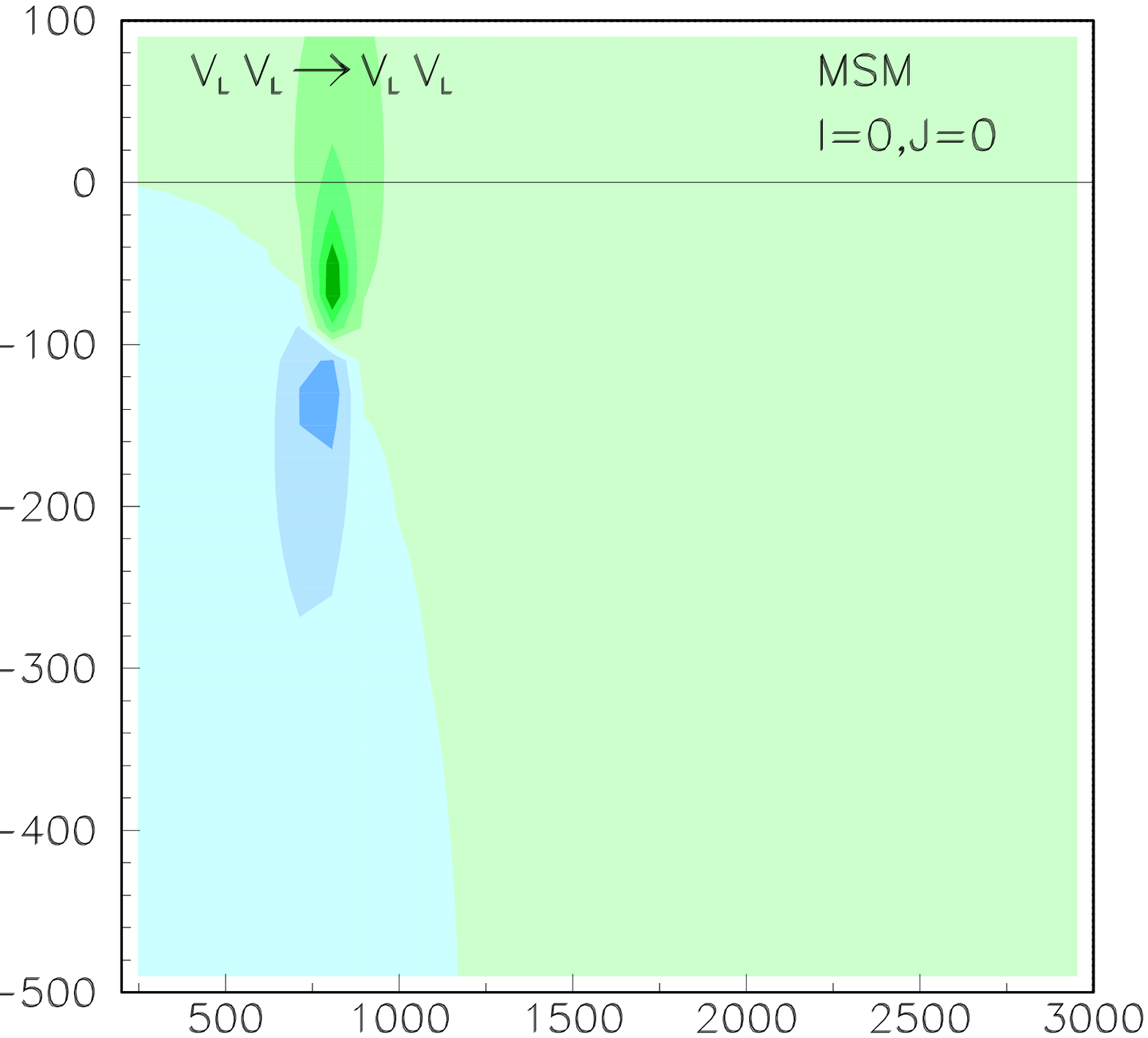}
\epsfysize=6.cm\epsffile{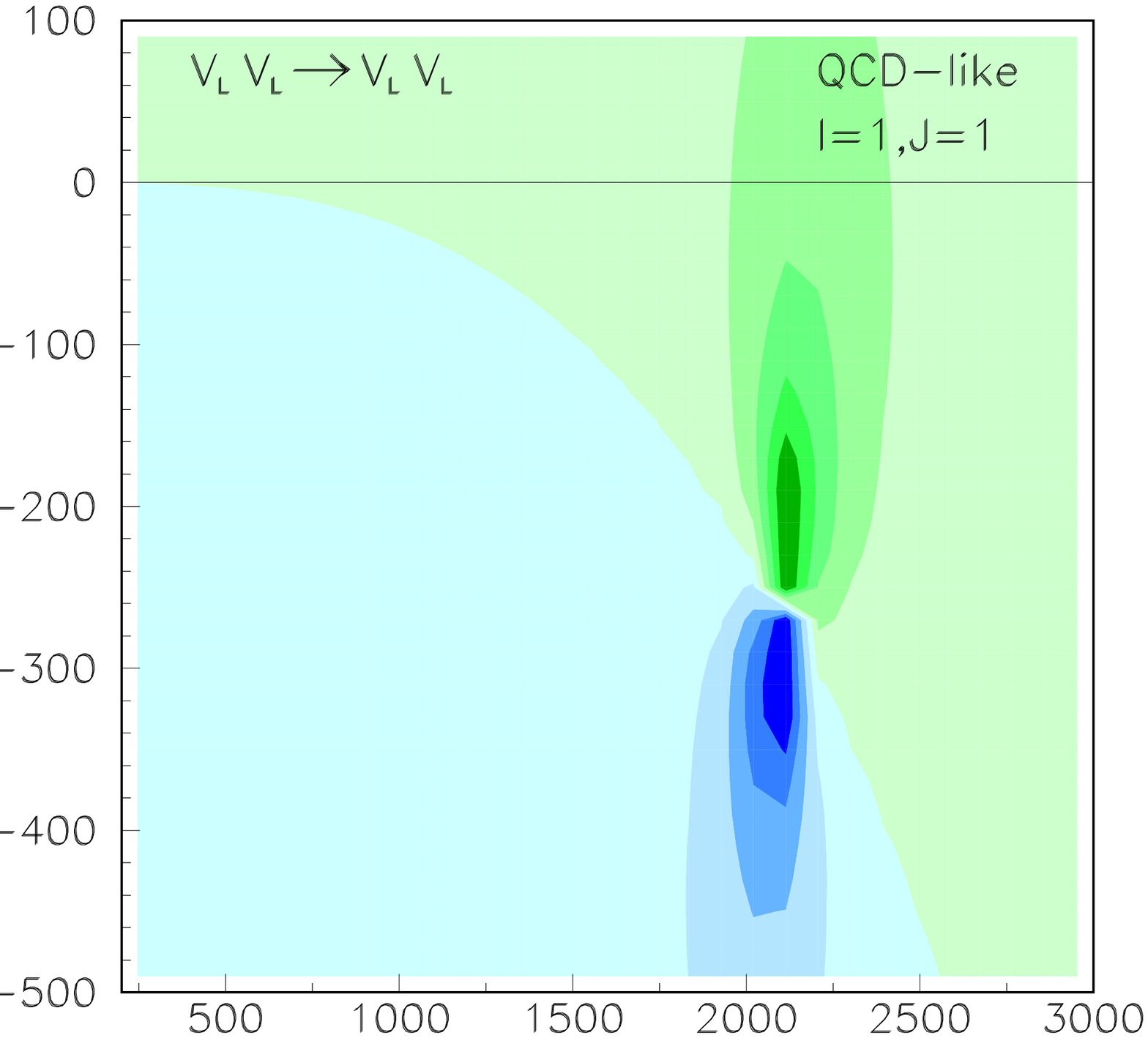}}
\end{center}

\vskip -.5cm

{\footnotesize
{\bf Figure 2.-} Contour plots of the Imaginary part
of the $V_LV_L\rightarrow V_LV_L$ chiral amplitudes.
It has been extended continuously through the cut.
Thus, above the real axis (straight line) is the first
Riemann sheet, and below the second. a) Pole of the scalar 
resonance in the $M_H=1 \mbox{TeV}$ MSM. b) Pole of the $\rho$-like
resonance in the QCD-like channel.}
\vspace{.5cm}
%----------END FIGURE 2

But it could well happen that there is a 
big phase shift background without a nearby pole.
Then the phase shift can cross $90^o$ and saturate unitarity but
we will not see the sudden increase in the phase shift.
That we will call "saturation".
As a matter of fact such big background phases are also produced by poles,
but they are very far away from the real axis. Then it is 
either possible to say that there is no resonance or
a very broad one. That is for instance the case of the $(I,J)=(0,0)$
channel in $\pi\pi$-scattering.
That channel has a huge enhancement in the phase shift that 
grows very rapidly at small energies 
(see Figure 1, which is a rescaled version).
Such an enhancement has sometimes been interpreted as a resonance:
the $\sigma$ particle. We will not address the $\sigma$ problem here.
The only thing that is more or less clear is that
such rapid enhancement should be produced by a pole \cite{Zou}
which is not very close to the real axis. Such a pole has been found
using the IAM and ChPT in approximately the correct position \cite{IAM}.

The position of the poles in our amplitudes does obviously depend
on the chiral parameters. Thus by varying $L_1$ and $L_2$ we can move the pole
far away from the real axis and create such saturation effects.
In Figure 3 in can be seen an example of that situation.
Following the discussion above, 
the pole is much farther away from the real axis than those in
Figure 2. As a consequence, the Breit-Wigner relations between 
its position and the physical parameters of an hypothetical resonance,
do not longer hold.
Notice also that the pole has changed its orientation. 

\leavevmode
\begin{center}
\mbox{
\epsfysize=6.cm\epsffile{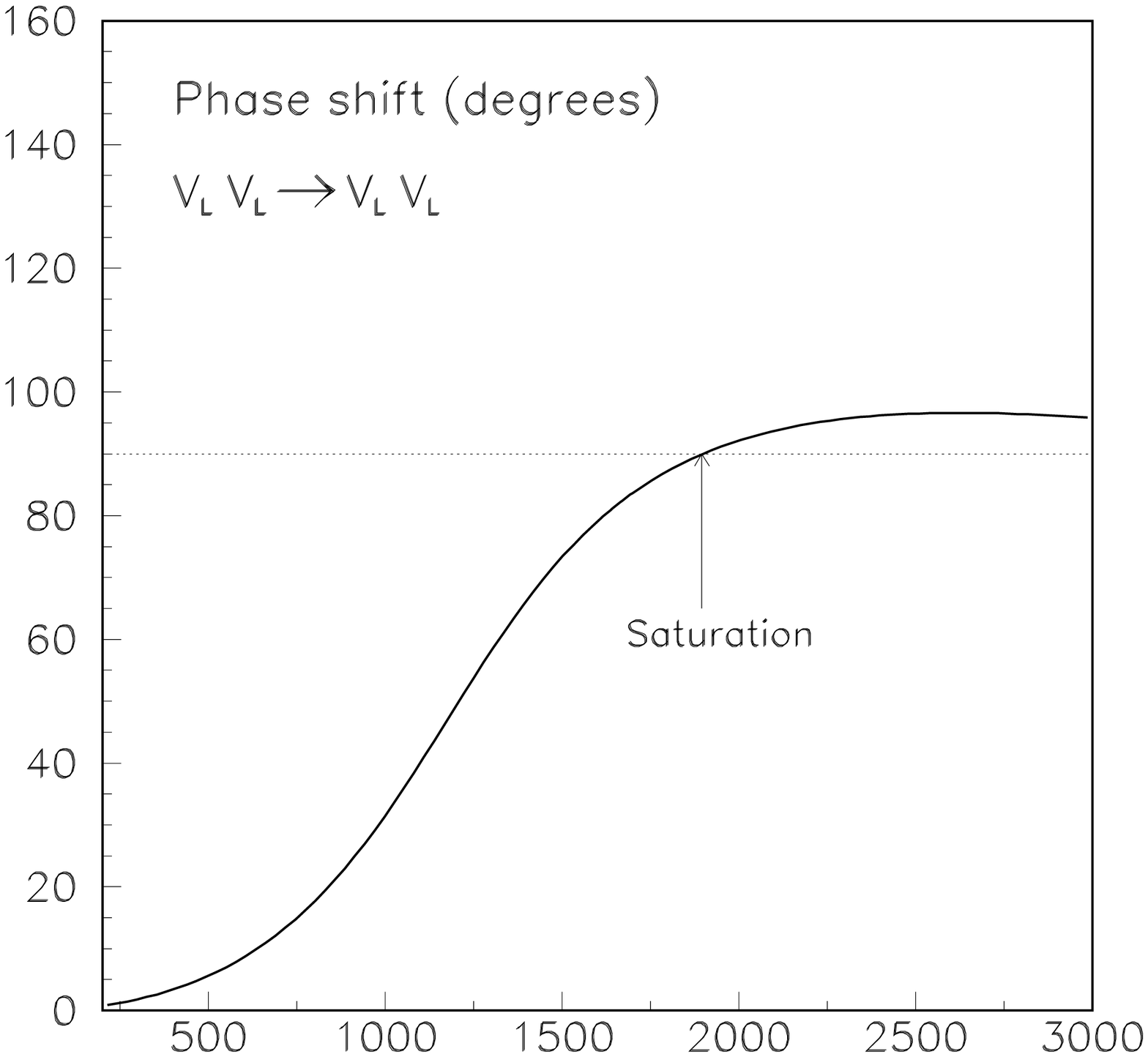}
\epsfysize=6.cm\epsffile{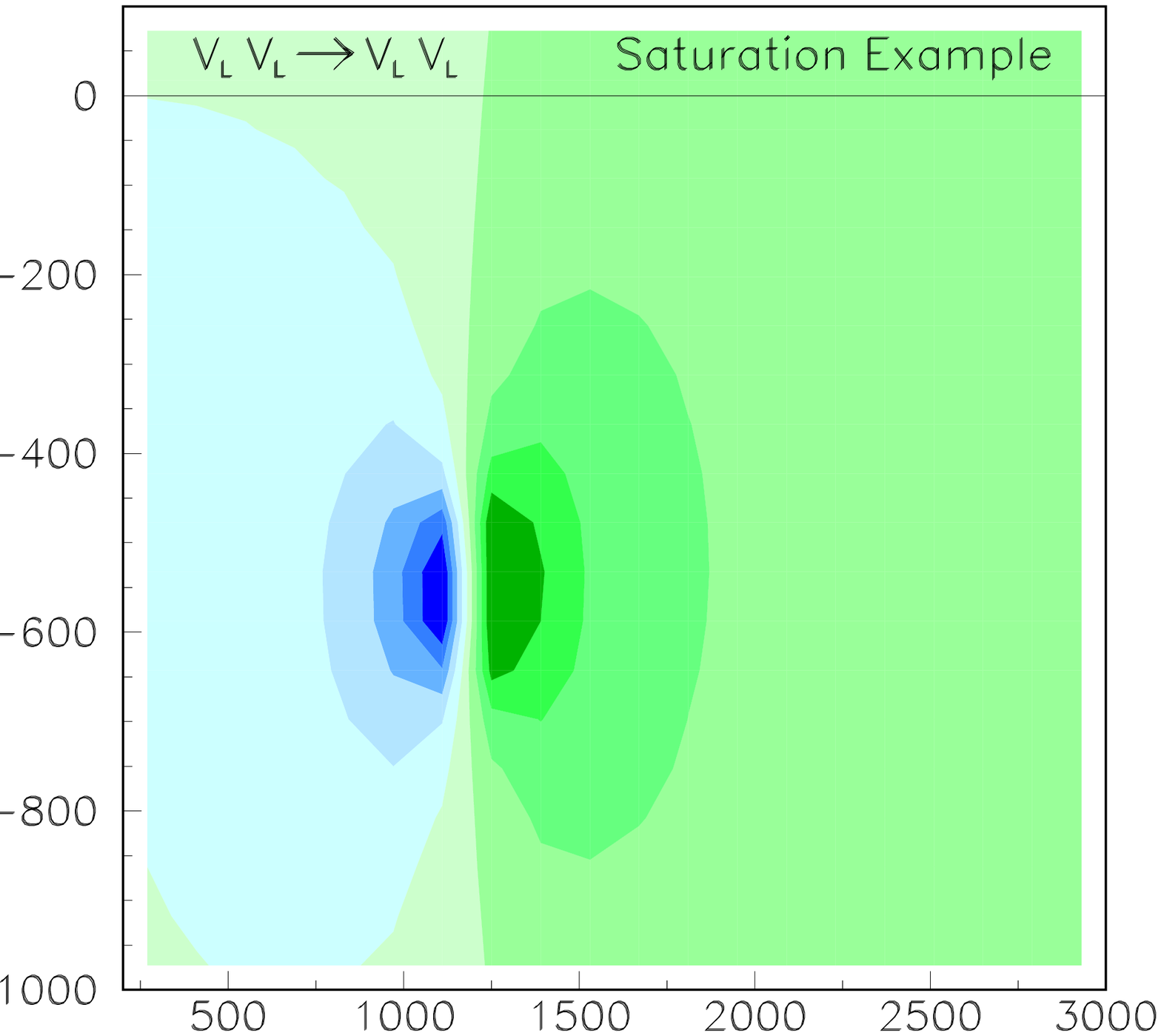}}
\end{center}
\vspace{ -.5cm}
{\footnotesize
{\bf Figure 3.-} 
On the left we show the phase shift of a channel
displaying a "saturation" effect. For the same model
we show on the right the contour plots of the Imaginary part
of the amplitudes.
Notice the change of the scale with respect to Figure 2. 
Observe that the pole
is far away from the axis and has changed the orientation too.}

\subsubsection{resonances in parameter space}
We have seen that the IAM and the chiral formalism yields
reasonable results in both reference models. Not only in terms
of resonances but also in non resonant channels. We have also shown
how the different features are described accordingly to the requirements
of analyticity and dispersion theory. Let us then explore the chiral parameter
space in order to get a qualitative description of the possible EWSBS.

In Figure 4 we show different contour plots in the $L_1,L_2$ plane.
We display the $10^{-2}$ to $10^{-3}$ range, since generically we expect the 
parameters to be of that order.

The contour plots have been obtained from the
calculation of the phase shifts in a $60\times60$ grid.
Using these phase shifts,
we have extracted two parameters: $M$, which is the energy
 at which $\delta_{IJ}=90^o$ and
\begin{equation}
\Gamma\simeq\left(M\frac{d\delta(s)}{ds}\right)^{-1}
\label{Gamma}
\end{equation}
The interpretation of these parameters 
has to be made carefully.
When $\Gamma\ll M$ they correspond to the mass
and the width of a resonance in the
Breit-Wigner approximation. Otherwise, the situation
is similar to our previous "saturation" example and
$M$ is just the point
where the amplitudes saturate unitarity.
In such case, $\Gamma$ should 
not be interpreted as the width of a particle, although
the saturation shape is broader for bigger $\Gamma$.
In addition, $M$ and $\Gamma$ are not related to the
pole position as in the Breit-Wigner formula.
Remember from Fig.3 that the pole not only moves away
from the real axis, but it also changes its orientation.

\leavevmode
\begin{center}
\mbox{
\epsfysize=5.8cm\epsffile{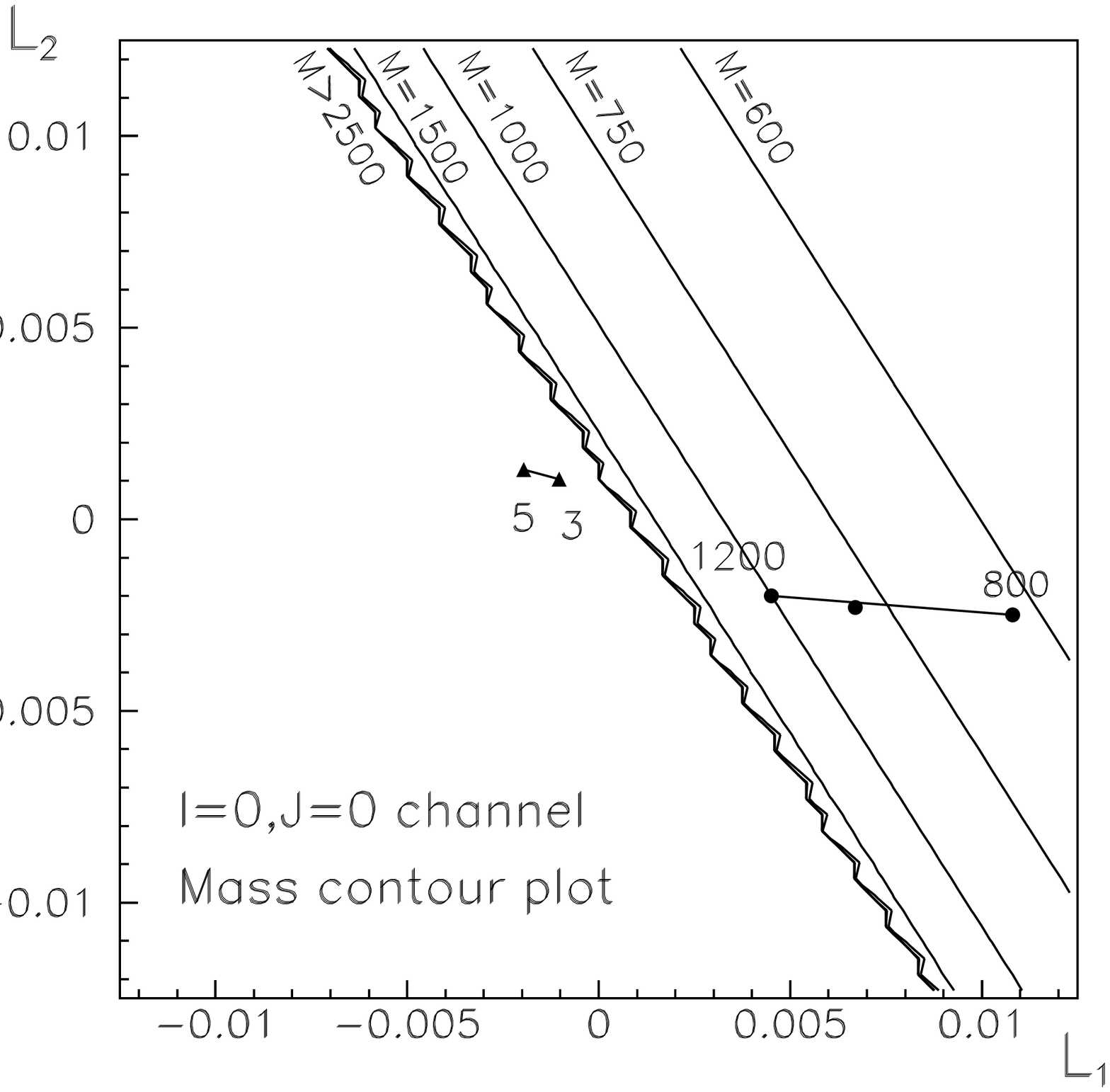}
\hspace{-4mm}
\epsfysize=5.8cm\epsffile{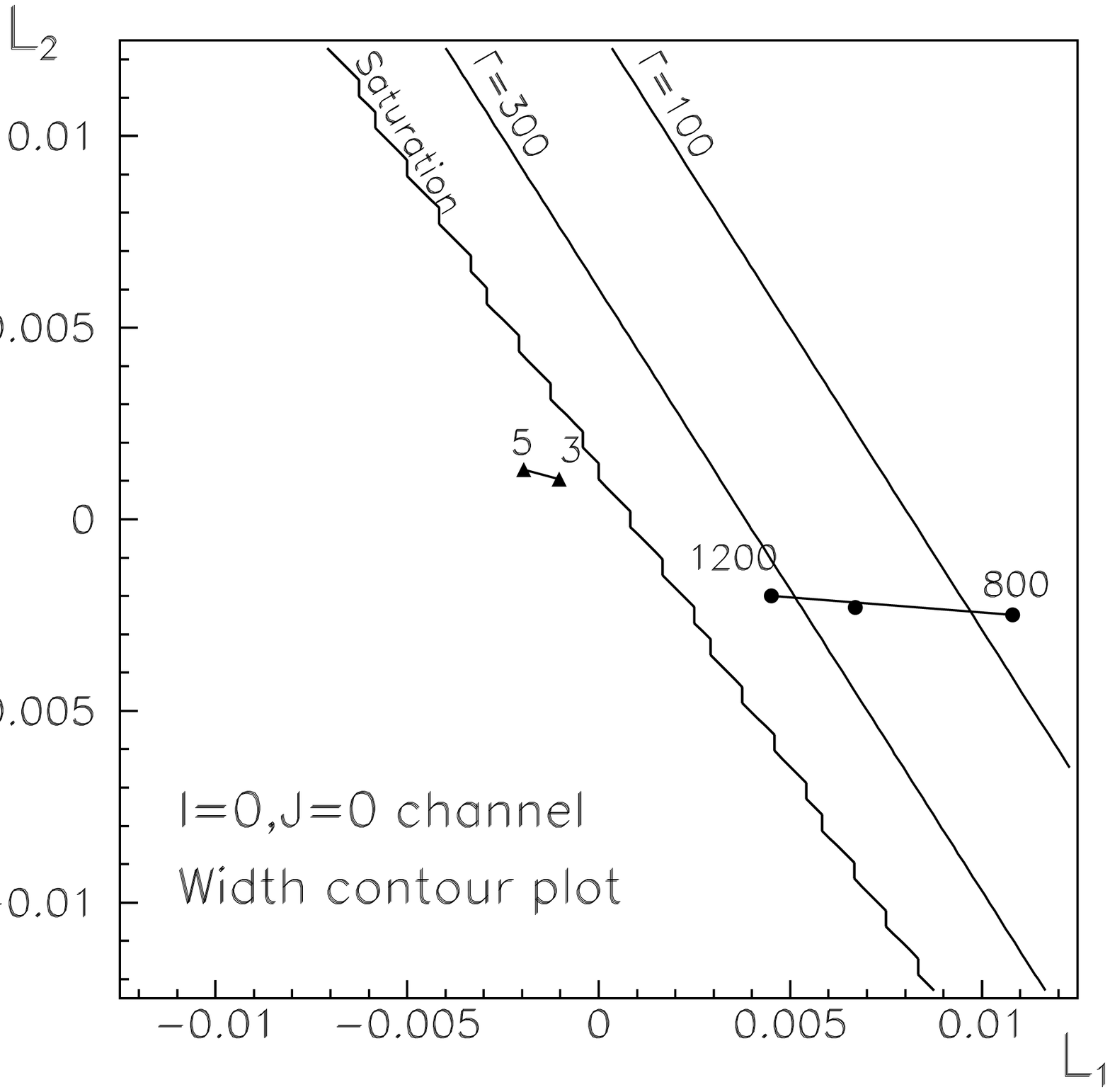}
\hspace{-4mm}
\epsfysize=5.8cm\epsffile{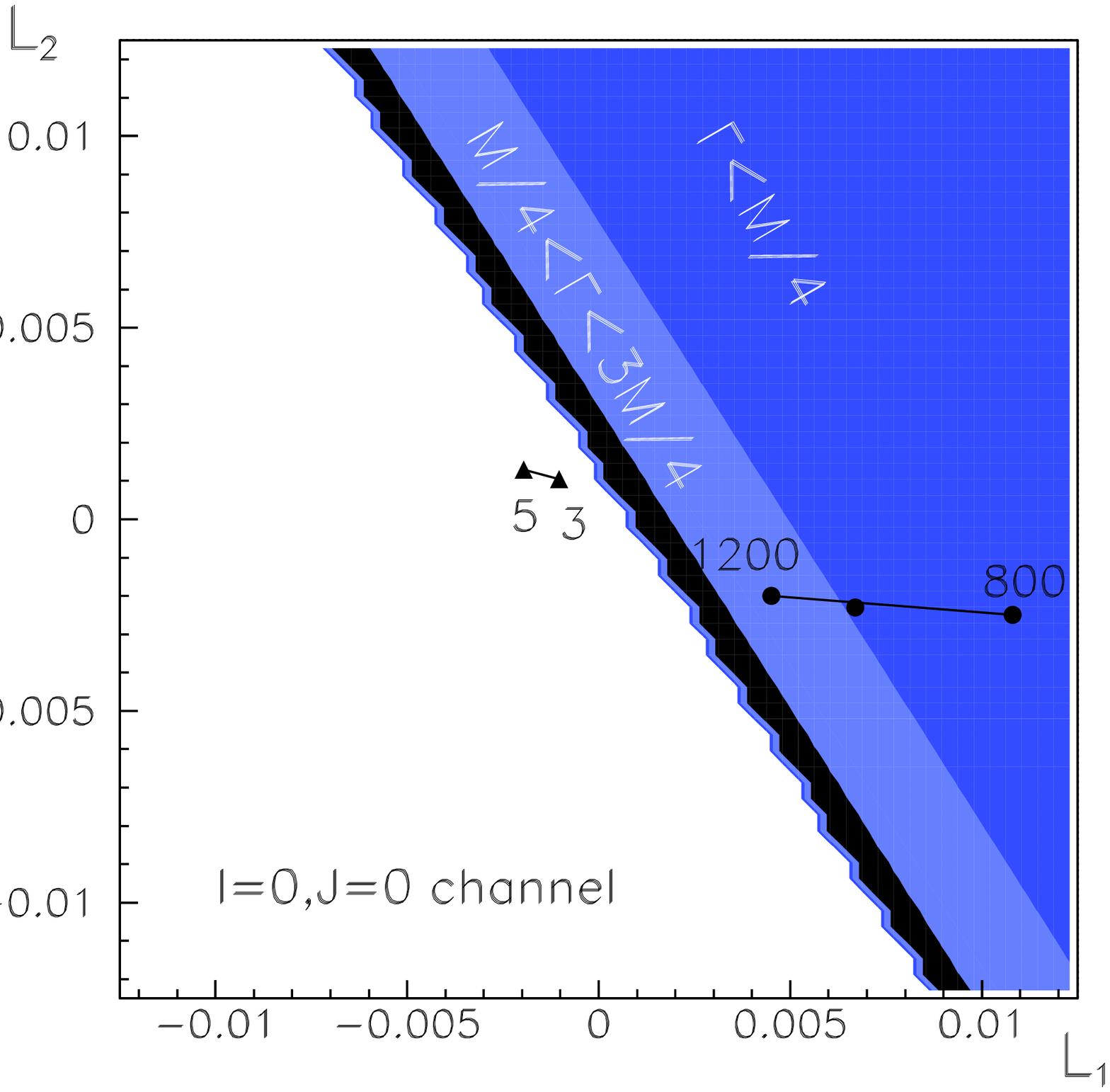}}
\mbox{
\epsfysize=5.8cm\epsffile{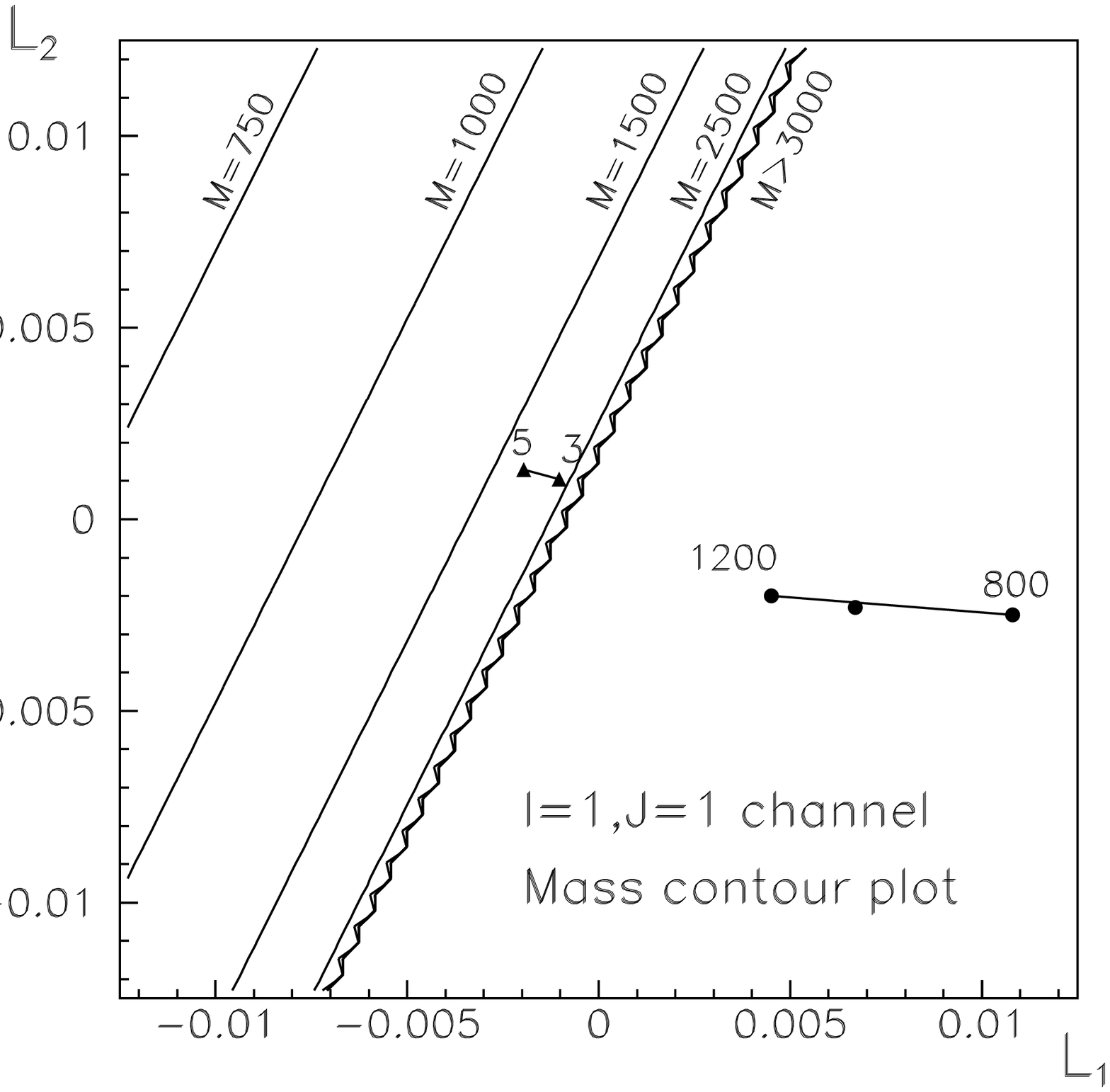}
\hspace{-4mm}
\epsfysize=5.8cm\epsffile{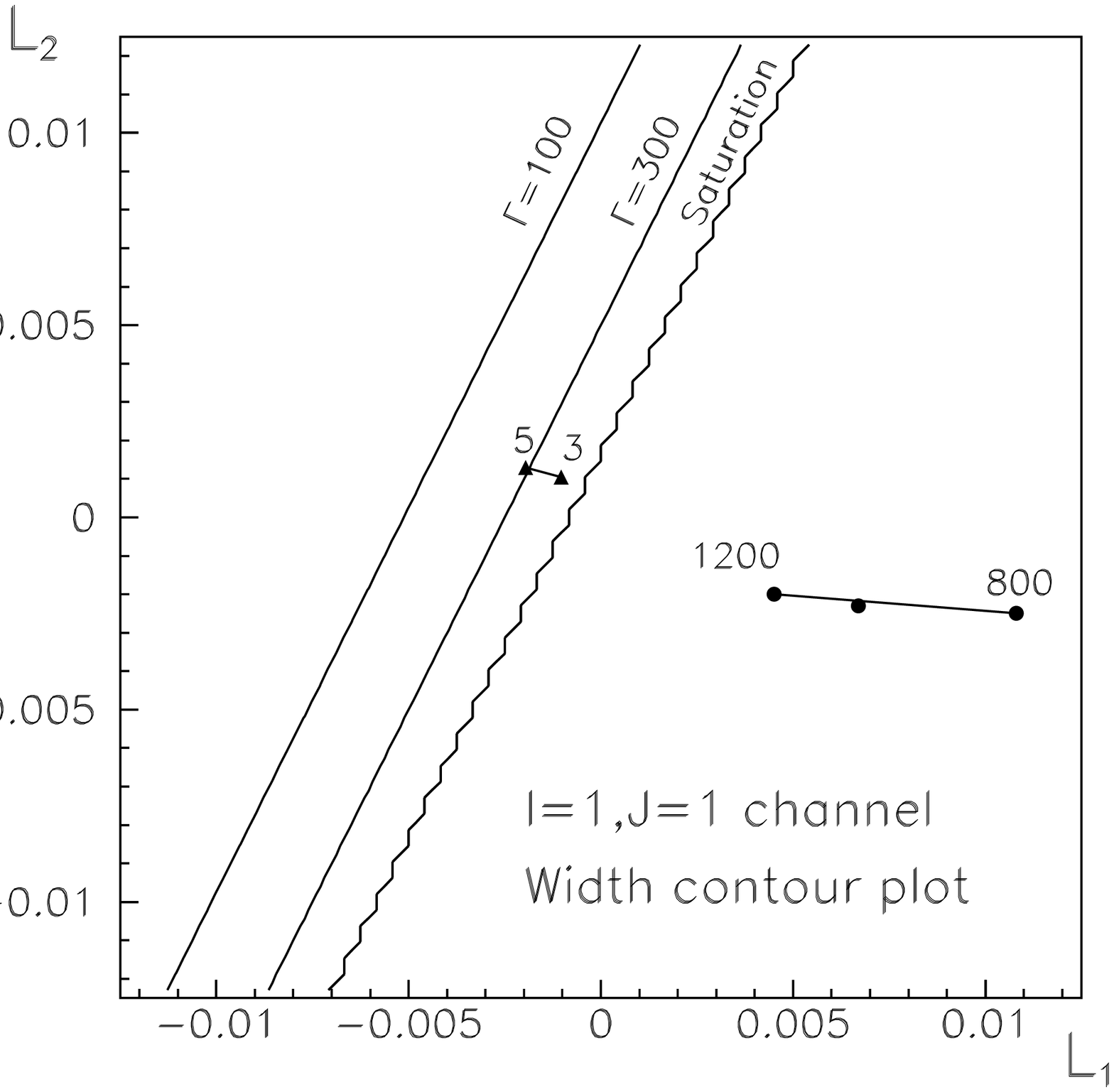}
\hspace{-4mm}
\epsfysize=5.8cm\epsffile{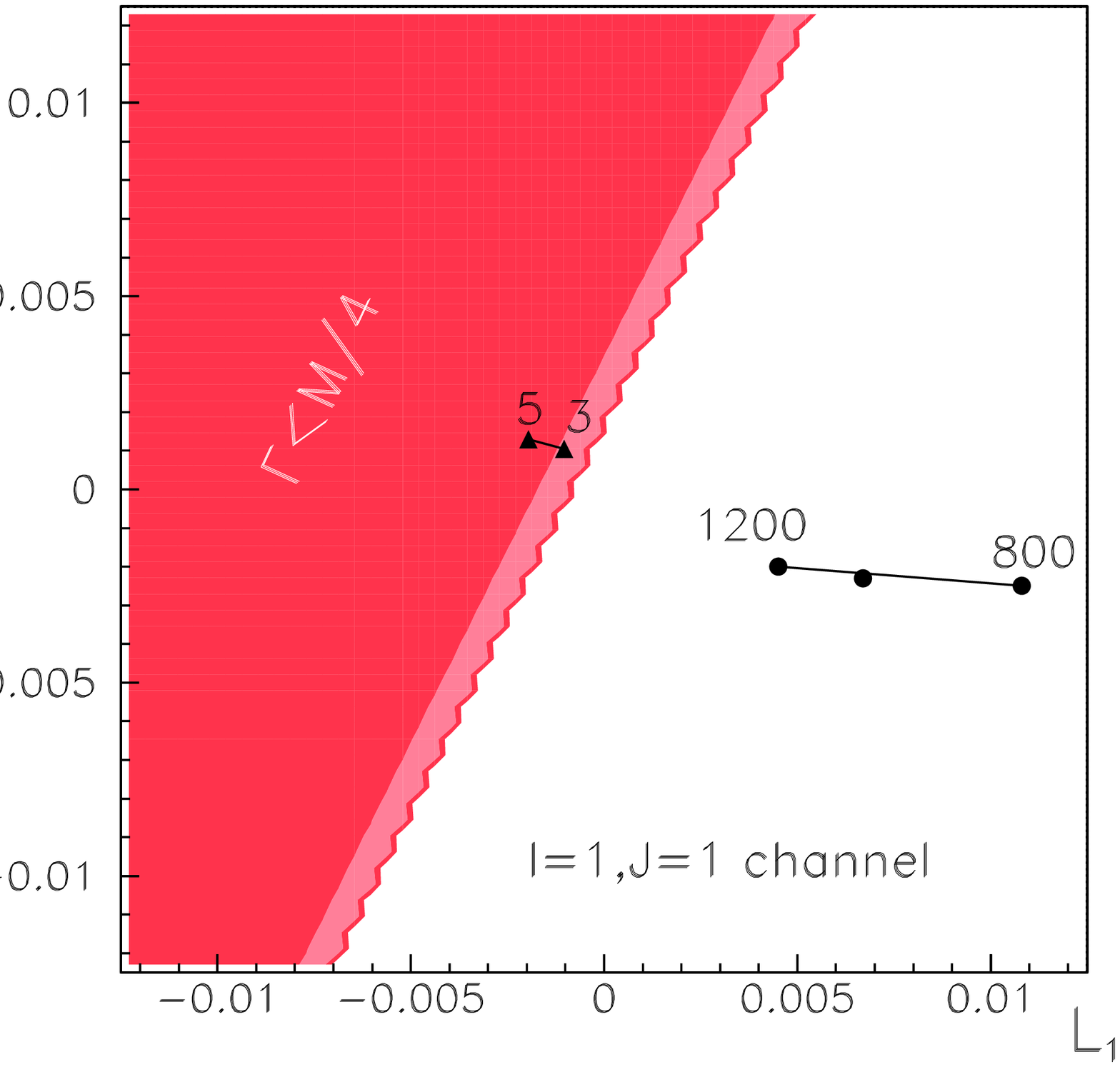}
}
\end{center}
\vspace{ -.5cm}
{\footnotesize
{\bf Figure 4.-} Plots in the $L_1,L_2$ plane for
the $(I,J)=(0,0)$ 
and $(1,1)$ channels.
The plots on the left give contour levels
of $M$. Those on the center give $\Gamma$.
The dark grey areas stand for narrow 
($\Gamma<4M$) resonances. The light grey areas for broad ($M/4<\Gamma<3M/4$)
resonances and the black areas for saturation.
White is no resonance or saturation below 3 TeV.
The black circles stand at the values of $L_i$ that mimic a MSM
with $M_H=800,1000$ and $1200 \mbox{GeV}$. The black triangles
represent QCD-like  models with 3 or 5 colors.}
\vspace{.5cm}

We are showing three plots for 
the $(I,J)=(0,0)$ and $(1,1)$ channels separately.
The contour plot on the left shows the values
of $M$. That on the center is a 
contour plot of $\Gamma$. 
In order to clarify
the meaning of these parameters, but also to get a qualitative
picture of the many possible strong scenarios, we have added a third
plot on the right. The dark gray area corresponds to "narrow" resonances.
For illustrative purposes, we define narrow as $\Gamma<M/4$. 
Roughly, this is what it is usually understood by a resonance.
Indeed, in QCD both the $\rho(770)$ and $K^*(892)$ satisfy this criterion.
The light gray area stands at those $L_1,L_2$ values where 
we get a broad resonance. In this case, broad means $\Gamma>M/4$ but
even though the width is not very small, it is still possible to
describe it with a pole and a Breit-Wigner.
Obviously, if we make even bigger the $\Gamma/M$ ratio
the Breit-Wigner description is no longer valid. That happens
more or less at about $\Gamma>3M/4$ and at those points
the black area starts, pointing the existence of a saturation effect.

\subsection{The I=2 channel}

We have already reviewed how the IAM in the chiral approach 
is able to reproduce the $I=0$ or $I=1$  resonances of
our reference models. In the literature, there have also
been proposed models with $I=2$ resonances 
(see \cite{Georgi} or \cite{Chivukula} and references therein). 
However, they do not correspond to the kind of models
that we are dealing with, since they always present
light resonances or states. Indeed, 
in \cite{Georgi} two models were built 
with $I=2$ resonances, one of them with
elementary and the other with composite doubly charged states.
In both cases their masses are $M_{++}\lappeq160\mbox{GeV}$ and
there are also single charge states with $M_{+}\simeq100\mbox{GeV}$.

Even more, the authors in \cite{Chivukula} 
slightly modified the MSM including an
$I=2$ resonance. Using tree level unitarity, they found
that the model does not make sense if its mass is bigger than 
$\sim 375\mbox{GeV}$. That bound becomes even smaller
as the scalar Higgs-like resonance gets heavier.
In the literature there are no models with an $I=2$ 
resonance and without light modes at the same time,
that respect the custodial symmetry.

Within our approach, we find a similar result but for
the general case. As soon as an $I=2$ resonant shape appears
in the spectrum, the models do not make sense. Indeed,
they present poles in the first Riemann sheet, within
the IAM applicability region. 

\subsubsection{Poles in the first sheet}

The $(I,J)=(2,0)$ phase shift is negative and
that can give rise to several problems related
to causality. In fact, saturation can also occur at
$\delta_{20}=-90^o$. However, if we apply
blindly Eq.\ref{Gamma}, we get a negative value.
Thus, even when $\vert \Gamma\vert \ll M$ we 
cannot say that there is a resonance, since its width would be negative.
From the analytical point of view, that situation corresponds
to a pole in the $1^{st}$ Riemann sheet, which is
forbidden.

As a matter of fact, 
the IAM yields poles in the first Riemann sheet
of the $(I,J)=(2,0)$ amplitude.
For instance,
it is possible to find poles in $t_{20}$ in the $1^{st}$ 
Riemann sheet at
about $\sqrt{s_{pole}}\sim 3300+i1750$ 
and $\sqrt{s_{pole}}\sim 4700+i7000$ for the
QCD-like and MSM parameters of Table 1.
However,  in the chiral approach
we are only allowed to use the IAM for energies
$\sqrt{s}\lappeq 4\pi v\sim 3\mbox{TeV}$.
We should not worry about the IAM results outside that region,
since it is not a good approximation there.
These poles are well outside a circle of that radius in
the complex plane and are not real predictions of the approach.
 In addition, when looking at
pion physics, the description of $\delta_{20}$ 
is correct with the IAM and qualitatively wrong (at high energies)
with plain ChPT \cite{Pade2,IAM}.

The problem is that the position of those
poles depends on $L_1$ and $L_2$. In fact, it is possible to
bring them close to the real axis and then the amplitudes
do not have a physical meaning.

Let us now recall that the chiral 
lagrangian does not meet all the requirements of a 
relativistic QFT. It respects
hermiticity, its amplitudes present a cut and an analytic
structure, etc... but it is not renormalizable.
It could well happen that, given a set of chiral parameters,
there is no underlying theory  
consistent with all the QFT requirements.
That could be enough to 
yield poles in the $1^{st}$ Riemann sheet. 
If we were able to develop a method to detect those poles, we
could rule out that parameter set as unphysical.
In the appendix we have shown that the IAM is able
to reproduce these poles when they are present
in the underlying theory.

The next step is to define how far these conflictive poles
should be to accept the IAM results. Looking at the MSM and QCD-like 
examples, we notice that they are not a problem if they lie
outside a $4\pi v= 3 \mbox{TeV}$ circle in the complex plane.
However, the IAM is a good approximation only near the
real axis and thus the above criterion 
could be too strict. There is a much more intuitive criterion
in order to exclude some values of $L_1$ and $L_2$.

\subsubsection{Wigner bound}

Indeed, there is a lower bound on the
phase shift derivative due to Wigner \cite{Wigner}.
Roughly it can be understood as follows:
Phase shifts can be interpreted as the delay
of the outcoming wave with respect to the incoming one.
When it is negative, the outcoming signal is {\em advanced}.
But that advance cannot be arbitrarily big. In the
classical case, $d\delta/dk>-D$, where $D$ is
the radius of the scatterer and $k$ the momentum
of the incoming particle. The wave nature of particles
does allow for a small violation of the previous equation.
Near a resonance,
it can be shown that $d\delta/dk>-(D+1/2k)$
\cite{Wigner}. For a general potential the definition of $D$
is not so evident, but intuitively it has to be related to its
 effective size or range. Notice that this bound is valid 
for the elastic case.

Let us then translate the above arguments to our problem.
First, in $V_LV_L$ scattering we are interested in the CM frame,
where the momentum is $q^2=s/4-m_W^2$.
Second, we have been using the $\Gamma$  parameter 
instead of the slope. Using Eq.\ref{Gamma}
our previous bound, in the CM, reads
\begin{equation}
\vert \Gamma\vert > \frac{2m}{M}\frac{1}{D+\frac{m}{M^2\sqrt{1-4m^2/M^2}}}
\simeq\frac{2}{M/(8\pi v^2)+\frac{1}{M\sqrt{1-4m^2/M^2}}}
\label{cota}
\end{equation}
where in the last step we have used as $D$ the scattering length
of the $t_{20}$ wave, which is the one we are interested in.
It seems to be a reasonable estimate of the effective size
of the potential. 
We will have to check that our results respect this condition.
To start with, both reference models satisfy it. Let us now
see what happens for other $L_1,L_2$ values.

\subsubsection{The IAM results}

In Figure 5 it is shown the result
of applying the bound in Eq.\ref{cota} to the IAM
$t_{20}$ amplitude. The area in black represents the area excluded,
whereas the white area is no saturation
of unitarity. Notice that there is only a very narrow strip where 
the criterion is respected and saturation occurs.
In this band, colored in grey,
the saturation point $M$,
 is always reached above $M> 2150\mbox{GeV}$, with 
$\vert\Gamma\vert>1050$. Surprisingly, 
the allowed $M$ and $\Gamma$ values are outside a $3 \mbox{TeV}$
region. But that is again
the first naive criterion of $\vert s_{pole}\vert > 4\pi v$.
Thus, our allowed paremeters yield
amplitudes that satisfy both criteria at the same time.
In the cross section these $M$ and $\Gamma$
parameters would give a 
very broad shape of a resonance (although it cannot be interpreted as a
particle) or a saturation effect.

\leavevmode
\begin{center}
\mbox{
\epsfysize=5.8cm\epsffile{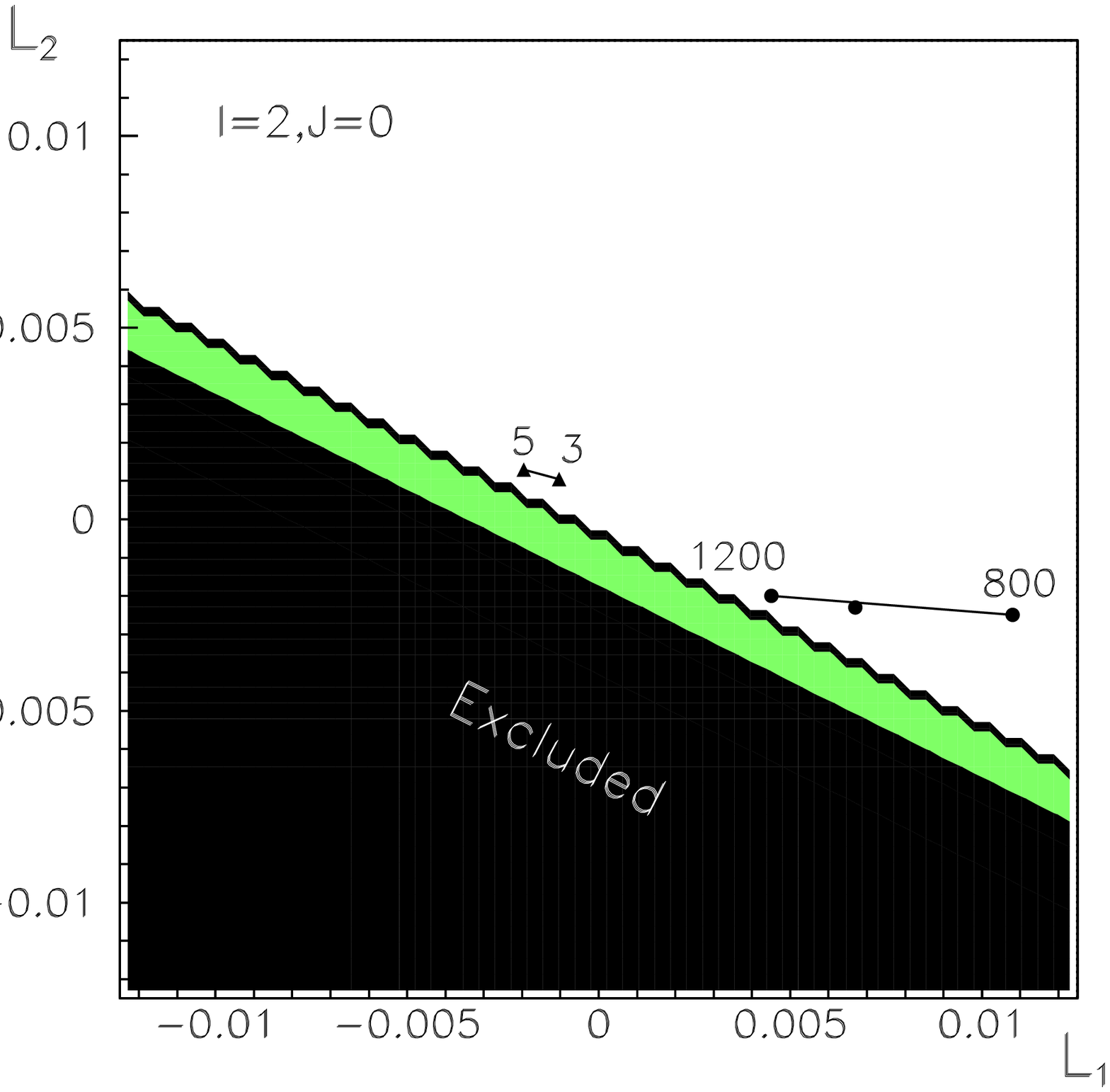}
}
\end{center}
\vspace{ -.5cm}
{\footnotesize
{\bf Figure 5.-} Contour plot in the $L_1, L_2$ plane.
The black area is excluded using the slope criterion and the IAM.
The shaded area indicates a broad saturation
shape in the cross-section. In the white area there is no saturation 
below 3 TeV.} 

\section{Discussion}

In the previous sections we have obtained the resonance spectrum
for the general strong EWSBS. Let us now review what is the physical
meaning of the results in the different $(I,J)$ channels:

\paragraph {\bf (0,0) Channel.} 
Concerning the MSM,
we have already remarked that the IAM yields 
a Higgs-like resonance.
As it can be seen in Figure 4, its mass is always 
smaller than $M_H$. As far as $M_H$ is the only relevant parameter,
for a given resonance mass there is a fixed value of the 
width. With respect to QCD-like models, we do not get any resonance,
but we get a considerable enhancement in this channel. 
This is the analogous of
the $\sigma$ particle problem in QCD. There is also a pole 
very far from the real axis and it does not saturate unitarity.

In the general case, once we fix $M$ we get 
a unique $\Gamma$ too, since this 
channel only depends on the $8L_1+5L_2$ combination.
That is not in conflict with
existing models where the mass and the width of a scalar resonance can be
adjusted \cite{Chivukula} independently.
 In those models, there are resonances
whose masses are $\Od(100)\mbox{GeV}$. In this work we are only 
studying those models {\em without} low lying resonances. In addition,
we have simplified the calculation to lowest order in $g$. When further
corrections are included, other $L_i$ come into play and different values
would change $\Gamma$. Nevertheless these effects are weaker and the
variations should be relatively small. 

Let us also notice that we can get narrow resonances, broad resonances
and that saturation occurs when $M\gappeq 1500$ (and then we cannot strictly 
speak of a mass).

\paragraph {\bf (1,1) Channel.} 
Again there is only one $\Gamma$ for every $M$,
since this channel only depends on $L_2-2L_1$. 
In contrast with the previous channel, 
we can see in Figure 4 that there are narrow resonances
up to $M\lappeq 2500$. The values where we obtain a broad resonance are limited
to a thin band, and we do not find what we have called "saturation"
below 3 TeV.

The IAM yields a clear resonance in the QCD-like models.  It is very narrow
although not as much as the real $\rho$. That is due to the fact that 
in QCD the GB (the pions) are proportionally more
massive than their analogous here (the $V_L$). 
It is also interesting to notice that vector like resonances
become lighter when we assume more technicolors.
As a consistency check, we do not get any resonance for the MSM.

\paragraph {\bf (2,0) Channel.} 

The interpretation of the results in this channel is more delicate.
The IAM is only expected to work near the elastic cut. 
It has been tested in pion physics \cite{IAM} and it yields the 
correct behavior in this channel. Nevertheless in ChPT
it presents poles in the $1^{st}$
Riemann sheet, although very far from the IAM applicability region.
They cannot be considered predictions of the approach.
However
the position of these poles depends on the chiral parameters,
and it is indeed possible to get them very near the axis.

At this point we should remember that the chiral formalism is not
renormalizable. It is not guaranteed that
for every value of $L_1$ and $L_2$ there should be an underlying
consistent theory. We have shown in the appendix that in case these
inconsistencies caused the appearance of a pole in the $1^{st}$ sheet,
and close to the unitarity cut,
the IAM should be able to reproduce it properly.
Consistently, when these poles are present we violate Wigner's
bound on the phase shift slope. This bound is respected when we 
take the poles very far away.

We therefore consider the existence of those poles and the violation 
of the Wigner bound as a strong hint
that the corresponding $L_1$ and $L_2$ are not allowed.
In Figure 5 we have shown the corresponding excluded region and
those values where we get a saturation of unitarity, which
always occurs at $M\gappeq 2000$. 
In any case these parameters should never be understood 
as those of a resonance.
Notice, once more, that the
$\Gamma$ parameter is fixed for a given $M$. 
That is due to the fact that this
channel only depends on $L_1+2L_2$.

The most striking consequence of this result is that there cannot
be heavy $I=2$ resonances unless some of our initial assumptions
are violated. Similar conclusions where found when trying to
build models with such $I=2$ resonances \cite{Georgi,Chivukula}:
it was not possible to make the $I=2$ resonances heavy unless
the other particles in the spectrum become very light. 
Even in that case, the $I=2$ resonances were never bigger than 
$\sim375\mbox{TeV}$.

\section{Conclusions}

In this work we have used the chiral lagrangian approach to
describe, with basically two parameters, the symmetry breaking
sector of the SM. Indeed, to any strong model respecting the custodial
symmetry and without light resonances, should correspond
a value of this two parameters. However, it is not
ensured that for any two parameters there should be an underlying 
consistent theory.
By means of the Inverse Amplitude Method,
we have scanned this two dimensional parameter space in search
for resonances or unitarity saturation effects.

We have reviewed how this approach is able to reproduce the
expected behavior of popular models like the Minimal SM or
a QCD-like model. Within the expected parameter range,
it is possible to find narrow resonances, broad resonances or 
simply saturation of unitarity in both the $I=0$ or $I=1$ 
weak isospin channels.
We have shown that the description of
these
resonances is consistent with the requirements of relativistic Quantum Field
Theory. Indeed, they are accompanied by poles in the second Riemann sheet
whose position is correctly related to the resonance mass and width.

Concerning the $I=2$ channel, we have found that imposing
elastic unitarity through the dispersive approach leads,
for some values of the parameters, to poles in the first Riemann sheet.
We consider
that as an strong hint excluding those values as unphysical.
As a consequence, it does not seem possible to find heavy $I=2$
resonances in models respecting the above assumptions.
That is in agreement with previous observations
concerning specific models with $I=2$ resonances.
Our result refers to the general strong scenario.
Nevertheless, it seems still possible to have very broad
shapes of unitarity saturation.

We have summarized the above results in Figure 6. We have colored the 
excluded area in black. The white areas are labelled according to their
unitarity features. There are two possible kinds of narrow resonances:
a Higgs-like (H) or a technirho ($\rho$). By narrow we mean 
that the width is less than one fourth of the mass. 
We have denoted a broader saturation shape in the $I$ channel,
by $S_I$. Notice that, in contrast to the most
popular models,
it is possible to have two narrow resonances, 
a resonance in one channel and saturation in another, or saturation in two
channels. Finally, the grey area correspond to those parameters
that do not saturate unitarity below 3 TeV. For those models
it is quite likely that the future colliders will not give
even a hint on the nature of the electroweak symmetry breaking sector.

\leavevmode
\begin{center}
\mbox{
\epsfysize=7.cm\epsffile{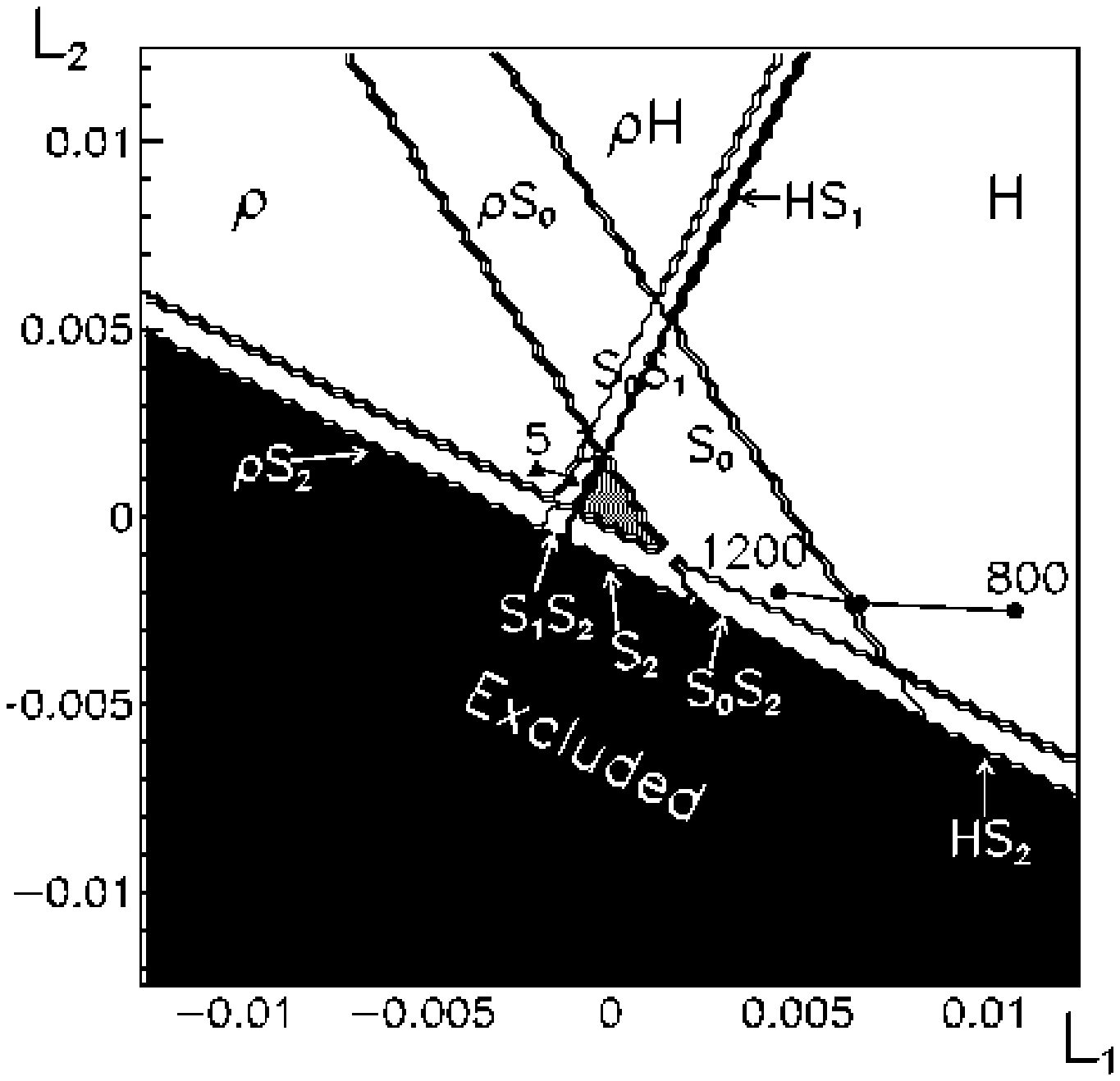}
}
\end{center}
\vspace{ -.5cm}
{\footnotesize
{\bf Figure 6.-} Resonance spectrum of the strong EWSBS in the $L_1,L_2$
plane. The black area is excluded. On the white areas, we have
represented broad resonances or saturation effects
in the $I$ channel by $S_I$; Higgs-like narrow resonances by H and
$\rho$-like narrow resonances by $\rho$. In the grey area there is
no saturation of unitarity, nor resonances, below 3 TeV.
The black dots represent the MSM with $M_H=800,
1000, 12000 \mbox{GeV}$ and the triangles a QCD-like model with 3 or 5
technicolors.}

\section*{Acknowledgments}

I would like to thank the Theory Group at Berkeley
for their kind hospitality and
the Jaime del Amo Foundation for a fellowship.
I have also greatly appreciated the patience of M.Suzuki 
during his explanations and 
discussions on the unphysical poles.
I am also indebted to R.S.Chivukula for his comments
on models with $I=2$ resonances as well as to J.J.Lozano
for helping me creating the figures.
This work has been partially supported by the Ministerio de
Educaci\'on y Ciencia (Spain) (CICYT  AEN93-0776). Partial support
by US DOE under contract DE-AC03-76SF00098 is gratefully acknowledged.

\appendix

\section{The derivation of the IAM}

In this appendix we will derive the IAM
method using dispersion relations.
Let us first remember that an elastic amplitude
has a left and right (or elastic) cut and thus two Riemann sheets.
A dispersion relation is nothing
but Cauchy's Theorem applied to one of these sheets. As a technical remark,
let us notice that our amplitudes are $\Od(p^4)\sim\Od(s^2)$.
Hence, we will have to divide by $s^3$ to ensure the vanishing
of the closing integral contour at $\infty$. That is, elastic chiral
amplitudes satisfy
\begin{equation}
t_{IJ}(s)=C_0+C_1s+C_2s^2+
\frac{s^3}\pi\int_{(M_{\alpha}+M_{\beta})^2}^{\infty}\frac{\Ima
t_{IJ}(s')ds'}{s'^3(s'-s-i\epsilon)} + LC(t_{IJ})
\label{disp}
\end{equation}
The $C_i$ subtraction constants can be determined from 
the chiral approach. 

Of course we only know how to calculate $t_{IJ}^{(0)}$ and
$t_{IJ}^{(1)}$ which is just a crude approximation to the above relations
\begin{eqnarray}
t_{IJ}^{(0)} &=& a_0+a_1s   \nonumber \\
t_{IJ}^{(1)} &=& b_0+b_1s+b_2s^2+
\frac{s^3}\pi\int_{(M_{\alpha}+M_{\beta})^2}^{\infty}\frac{\Ima
t_{IJ}^{(1)}(s')ds'}{s'^3(s'-s-i\epsilon)}+LC(t^{(1)}_{IJ})
\label{disp1}
\end{eqnarray}
Our aim is to obtain a much better description of the right cut.
That is because
resonances are understood as poles in the second Riemann sheet,
which is obtained continuously from the cut.

The relevant point is to realize
that the inverse amplitude can be calculated {\em exactly} on the
{\em elastic cut}. Indeed, using Eqs.\ref{uni} and \ref{pertuni}
we find {\em on the right cut}
\begin{equation}
\Ima \frac{t_{IJ}^{(0)2}}{t_{IJ}}
=-t_{IJ}^{(0)2}\frac{\Ima t_{IJ}}{\mid t_{IJ}\mid^2}=
-t_{IJ}^{(0)2}\sigma = -\Ima t_{IJ}^{(1)}
\end{equation}
Notice that we have normalized the inverse amplitude with the
real factor $t_{IJ}^{(0)2}$. Apart from the poles, this function
has the same analytic structure of $t_{IJ}$. 
Observe that the poles of $t_{IJ}$ are zeros of $G$ and viceversa.
Thus we can write
\begin{eqnarray}
\frac{t_{IJ}^{(0)2}}{t_{IJ}}&\simeq& a_0+a_1s-b_0-b_1s-b_2s^2 \nonumber  \\
&-&\frac{s^3}\pi\int_{(M_{\alpha}+M_{\beta})^2}^{\infty}\frac{\Ima
t_{IJ}^{(1)}(s')ds'}{s'^3(s'-s-i\epsilon)}-LC(t^{(1)}_{IJ})+PC(G)
\simeq t_{IJ}^{(0)}-t_{IJ}^{(1)}
\end{eqnarray}
where we have approximated $LC(G)\simeq LC(t^{(1)}_{IJ})$ and
we have neglected $PC(G)$. That is
\begin{equation}
t_{IJ}\simeq
\frac{t_{IJ}^{(0)2}}{
t_{IJ}^{(0)}-t_{IJ}^{(1)} }
\end{equation}
Which is the IAM method. 
In the text we have already commented
its advantages, but there are also some limitations:
\begin{itemize}
\item We have only used elastic unitarity, and
that limits the validity at high energies where 
the first two body {\em inelastic} threshold appears \cite{IAM}.
\item We have also 
neglected the pole contributions of $G$ and thus we are not able 
to describe  Adler zeros below threshold.
\item Finally, we have approximated the left cut
of the inverse function by that of $t^{(1)}$. 
Hence we violate crossing symmetry. In addition we only reproduce the
leading but not the subleading logarithms. 
\end{itemize}

Notice, however,
that the expansion of the IAM at low energies is again
the chiral expansion $t_{IJ}\sim t_{IJ}^{(0)}+t_{IJ}^{(1)}$
so that the error in this approximations is $\Od(s^3)$.
At higher energies, the contribution
from the left cut 
and poles below threshold become less relevant, dur to the
$(s'-s)$ factor in the deenominator.
 Their effect will be to change slightly
the position of the resonance. In previous applications to ChPT it has been
found that this shift is usually smaller than $15\%$ \cite{IAM}.
As far as we are only interested in a qualitative description
of resonances, they will be neglected. Very recently, however, it has been
proposed an improved version of the IAM \cite{Penn}, although it 
does not yield such a simple formula. That is why we will not use it here.

Finally, let us remark that 
we have only needed the dispersion relation for the
inverse amplitude as well as those for the {\em approximated}
amplitudes, which do not have poles.
Even if the theory is pathological and presents
poles in the first sheet, the IAM derivation is still valid.
These poles in the amplitude become zeros of the inverse amplitude
and they do not change the analytic structure.
We can thus use the very same expression of the IAM in Eq.\ref{IAM}
to detect poles in the $1^{st}$ Riemman sheet. 
However, we still have to remember that the 
approximations we have done limit the validity of the method
to a region close to the elastic cut. Any feature, including poles,
 outside that region does not deserve any consideration.

\end{document}